# Application of new conformal cooling layouts to the green injection molding of complex slender polymeric parts with high dimensional specifications

Abelardo Torres-Alba[1-2], Jorge Manuel Mercado-Colmenero[1-2], Juan de Dios Caballero–García[1-2] and Cristina Martin-Doñate[1-2*]

[1] Department of Engineering Graphics, Design and Projects, University of Jaen, Campus Las Lagunillas s/n. Building A3-210, 23071 Jaen, Spain; ata00001@red.ujaen.es (A.T.-A.); jmercado@ujaen.es (J.M.M.-C.); jdcg0004@red.ujaen.es (J.D.C.-G.)

[2] Research Group Ingdisig Jaen. University of Jaen, 23071 Jaen, Spain

\* Correspondence: cdonate@ujaen.es; Tel.: +34-953212821; Fax: +34-953212334

**Abstract:** Eliminating warpage in injection molded polymeric parts is one of the most important problems in the injection molding industry today. This situation is critical in geometries that are particularly susceptible to warping due to their geometric features, as occurs with topologies of great length and slenderness with high changes in thickness. These features are, in these special geometries, impossible to manufacture with traditional technologies meeting the dimensional and sustainable requirements of the industry. The paper presents an innovative green conformal cooling system that is specifically designed for parts with slender geometric shapes, highly susceptible to warping. Additionally, the work presented by the authors investigates the importance of using highly conductive inserts made of steel alloys in combination with the use of additively manufactured conformal channels in reducing influential parameters such as warpage, cooling time and residual stresses in the complex manufacturing of long and slender parts. The results for a real industrial case study indicate that the use of conformal cooling layouts decreases cycle time by 175.1 s - 66% below the current cooling time, the temperature gradient by 78.5% specifically 18.16 ºC, the residual stress by 39.78 MPa or – 81.88 %, and the warpage by 6.9 mm or- 90.5%. In this way, it is possible to achieve a final warping in the complex geometry studied of 0.72 mm under the maximum value required at the industrial level of 1 mm. The resulting values obtained by the researchers present a turning point from which the manufacturing and sustainability in injection molding of said plastic geometries is possible, taking into account that the geometric manufacturing features analyzed, will present a great demand in the coming years in the auto parts manufacturing industry

## 1. Introduction

The plastic manufacturing process of injection molding is currently considered one of the most widely used plastic transformation technologies worldwide [1]. It is estimated that 70% of consumer products incorporate components manufactured with this production process [2-3]. The global scope of the plastic injection molded market was estimated at USD 265.1 billion in the year 2020, expected to grow annually by 4.6% between 2021 and 2028[4].

Although the demand for plastic components comes from various industrial fields such as packaging, electrical, electronic products, and medical devices, the automotive industry is one of the areas that produce the largest number of plastic parts, with a demand valued at 45,000 million in 2020 estimating the growth of up to 65 million USD in 2027 mainly in Europe and China [5]. The automobile requires the design of multiple plastic parts such as wheel covers, casings, interior parts, complex optical parts, etc. This fact causes manufacturers to increase the use of injection molded parts due to their recyclability, durability, vibration control, strength, etc [6-8].

The threat of climate change is high on international, European, and national political agendas [9]. The commitments set out in the Paris Agreement set the baseline for cost-effective greenhouse gas emission reductions. To achieve climate neutrality in 2050, the EU raises the need to decarbonize the industry by increasing its competitiveness, demanding that technological products and processes be climate neutral [10]. The injection molding technology, processes around 117 million tons of thermoplastics worldwide every year, estimating the use of electricity in injection machines at approximately 105,109 Kw.h. resulting in 80 million metric tons of $CO_2$ emissions per year [11].

Technologically, plastic injection molding is based on the process of melting and subsequent solidification of the plastic material that has been injected at high pressure and temperature into the mold cavity [12-14]. Among the stages of the molding cycle, the cooling phase is the one with the highest energy and time consumption, being around 70% of the total cycle manufacturing time [15-17]. In a conventional molding process, water, acting as a coolant, flows continuously through a set of channels in the mold, starting the solidification process of the injected melt when it comes into contact with the surface of the cavity at a lower temperature [18-19].

Decreasing the molding cycle time is a key factor in improving energy efficiency and the sustainability of the process [20-21]. From the perspective of increasing productivity, an efficient cooling design is critical, since the cooling phase takes up a large part of the cycle time. Designing adaptative channel geometries capable of maintaining a constant distance between the part surface and the cooling channels, even in those difficult areas where material accumulation occurs, can dramatically decrease cooling time and increase the uniformity of temperatures at the surface of the molded part [22]. In the same way, a uniform distribution of temperatures on the surface of the part helps not only to reduce the cooling time but also to improve the quality of the part in terms of metrology [23]. An unbalanced temperature map can cause warpage in the plastic part and result in the rejection of manufactured parts, which goes against the industrial sustainability criteria currently required [24].

In industrial practice, the warpage of molded parts is an even bigger problem than shrinkage. Warpage is a consequence of unequal shrinkage values [25]. The injection molding process significantly affects the warping phenomenon in the geometry obtained due to non-uniformities in the mold wall [26]. Particularly of the parameters of manufacturing by molding, the surface temperature of the mold is the most important in the physical phenomenon of warping [27]. The differential cooling of the part occurs when the surfaces are at different temperatures, specifically in the hot spots of the mold, such as core pins, corners, thick areas, gates, etc. Hot spots cause problems by adding extra crystallinity to the material and prolonging the part's cooling time so that the area that cools the longest, shrinks the most [28].

The cooling of the mold in the traditional way is achieved by making straight channels in both cavities of the mold. Despite the economic manufacture of this type of channels, they are not useful in the uniform cooling of parts with complex geometry. The difficulty in establishing a uniform distance between the channel and the topology of the piece prevents this length from being kept constant, preventing the uniformity of temperatures on the surface of the piece and increasing the number of rejections in production. Additive manufacturing processes allow the manufacture of cooling channels following the contour of the part [29], giving design flexibility. Regarding the sustainability of the production process, additive manufacturing allows the use of shorter supply chains, as well as more efficient use of manufacturing materials. Likewise, conformal cooling channels reduce cycle time and energy consumption, making them green channels. Compared to traditional cooling channels, the use of conformal cooling channels decreases the cooling time to a maximal range of 15 to 50%. Beard et al. [30] decreased the cooling cycle time in a range of 20 to 40%. Torres et al. [31] reduce cycle time by 13% in very complex optical parts with a high thickness ratio. Sayfullah and Masood [32] reduced the total cycle time by 35% as well as the maximum temperature by 30%, improving the differential shrinkage and therefore its warpage. Mazur et al. [33] obtained improvements of 5º C in the maximum temperature, improving the shrinkage of the part as well as its warpage and Smidt et al. [35] improved the cooling time by 32.7% reducing the warpage by 28.2%.

Given the high influence of conformal cooling channels in reducing cycle time and improving part uniformity, different applications and designs have been proposed. Concerning the geometry of the layout, the spiral layout is one of the most used [34]. Unfortunately, these channels present sudden turns, increasing the pressure drop and slowing down the flow, weakening the effectiveness of the cooling process. To avoid these problems, other authors use layouts whose axes follow a zigzag [35-36] design. For highly complex geometries, it is possible to use cross-linked channel designs [37], porous [38], lattice [39-40] or vascularized [41] topologies. However, all these geometries [35-41] have the problem of requiring a connection between channels, so in the event of an obstruction caused by foreign elements, it could not be easily removed from the layout. Another way to cool complex cores is the use of inserts made of high thermal conductive materials such as high conductive steel alloys. The high thermal conductivity of these materials helps in the process of reducing the cycle time while maintaining the corrosion and oxidation resistance of the mold [42].

Reducing the warpage of a plastic part to a minimum value is a highly significant factor, especially for complex optical parts where dimensional accuracy affects the functionality of the vehicle [43-45]. The current designs demanded in the industry and especially in the automotive sector, increasingly complicate the manufacturing process by molding. The trends place more and more emphasis on the development of the visual or aesthetic appearance of the vehicle, complicating the functional aspect of the molded parts. This fact increases if we take into account that, in addition to style, some of these pieces have to fulfill an optical function at the level of lighting or light signalling by the standards. The style implemented today in the automotive industry highlights futuristic concepts characterized by large-format stylized details that incorporate new technologies such as LEDs. These technologies require a marked precision in the dimensional aspect of the pieces, so it is necessary to look for designs that improve the current conception of the injection molding process. The procedures and tools used to date do not guarantee an optimal result in terms of the aesthetic or functional aspect of the piece, as well as of the manufacturing process that makes the activity competitive.

The use of long plastic parts with geometric slenderness including areas with large thickness ratios is becoming more and more common in the automotive field and more specifically in the area of optical plastic parts using LED technology. The upcoming lines of aesthetics and functionality that will come in the coming years for the car demand a type of lighting not only external but also internal to the vehicle. These optical plastic parts require high aesthetic and dimensional specifications, being also characterized by a great length and slenderness. These geometries cannot be manufactured with quality using the current means of production. To avoid these problems, the use of conformal cooling layouts together with high thermal conductivity materials can help to obtain a uniform and balanced temperature map in these complex parts. Unfortunately, currently, the application of green conformal cooling channels for the cooling of injection molds is a complex procedure from the initial design to the final manufacturing process of the mold. Despite previous literature and relevant review articles, unfortunately, there are still significant gaps in the research and application of conformal cooling layouts [22].

To avoid the problems raised so far, the paper proposes a new industrially sustainable conformal cooling system that can reduce cycle times, longitudinal warpage, and re-sidual stresses in the manufacturing of highly complex injection molded parts, characterized by their great length and slenderness as well as thick points. Research has not yet been conducted at a scientific or industrial level on the application of conformal layouts to the manufacture of this type of plastic geometry that is impossible to manufacture meeting the industry's requirements using traditional methods. The results obtained by the authors mark a turning point in the manufacture by molding of this type of topology, given that the geometric manufacturing features studied in the paper will present a very high demand in the coming years in the automotive industry [46-47]. The research results obtained show that the use of conformal cooling layouts in the cooling of injection molds makes it possible to manufacture parts following the high dimensional requirements demanded by current industries. Finally, the results presented by the authors are fully in line with the sustainability and energy-saving criteria demanded by the European Union and the rest of the world in terms of the manufacture of industrial plastic parts.

## 2. Materials and Methods

*2.1 Geometric analysis and functional description of the plastic part manufactured via injection molding*

The analysis and determination of the geometric, technological, and functional requirements of the plastic part directly influence the process of geometric design and dimensioning of the main elements that make up the mechanical systems of the injection mold. Likewise, geometric aspects such as the complexity of the surfaces, the distribution of the thickness map, general dimensions, and geometric tolerances, not only determine the final configuration of the injection mold but also establish the technological parameters that control its correct manufacturing process. That is why, in this section, the main geometric features and parameters of the plastic part under study are described in detail, as well as the definition of its main application and functional requirements.

The plastic part object of study is defined as a chimsel. As shown in Fig. 1, this real case study is an industrial part belonging to a LED - collimator lighting system, used in the automotive industry. The main function of this optical part is to collect the light coming from the LEDs using several collimating elements (see Fig. 1) and project it onto a light output surface, geometrically composed of a large number of small squares with closed angles "Pillows" in charge of distributing the light according to the desired purpose (see Fig. 1). In this way, the light generated by the LED lighting elements is directed along the plastic part according to the functionality and standards established by the automotive industry.

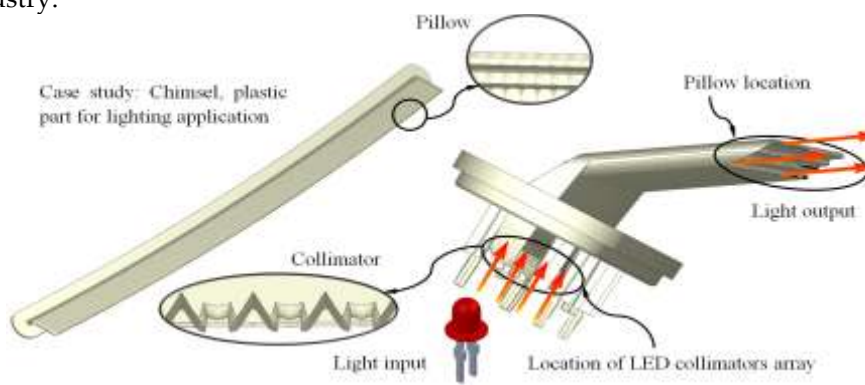

**Fig. 1** Description of the plastic part for lighting application case study

It should be noted that, as it is an optical part, the dimensional tolerances required to meet the requirements and standards of functionality and optical criteria are more demanding, relevant, and significant than for any other plastic part. In general, the final or resulting dimensional tolerances in plastic geometries are closely related to the design of the main components of the injection mold. This is especially true for the elements that make up the cooling system. Similarly, it is equally pertinent to establish the technological parameters that define the cooling phase within the cycle of manufacturing the plastic part. After manufacturing, warping can be generated due to physical concepts related to the cooling phase, including distribution of the cooling temperature gradients on the surface of the plastic part, residual stresses caused by a nonuniform cooling temperature, and the associated unbalanced volumetric shrinkage of the thermoplastic material. Therefore, an optimal design and dimensioning would favor the reduction of the final warpage of the plastic part, and therefore, achieve its functional requirements.

During the cooling phase, the most relevant features and geometric parameters are the thickness map and the main part dimensions (width, length, height), see Fig. 2 and Fig. 3, being the resulting variables the time needed until reaching the ejection temperature, the distribution of volumetric shrinkage, and the warpage after the cooling phase. These parameters mainly determine the economic viability, the efficiency, and sustainability of the manufacturing process as well as the compliance with the dimensional tolerances and the functional requirements of the plastic part. Table 1 shows the magnitude of the main geometric parameters of the plastic part under study.

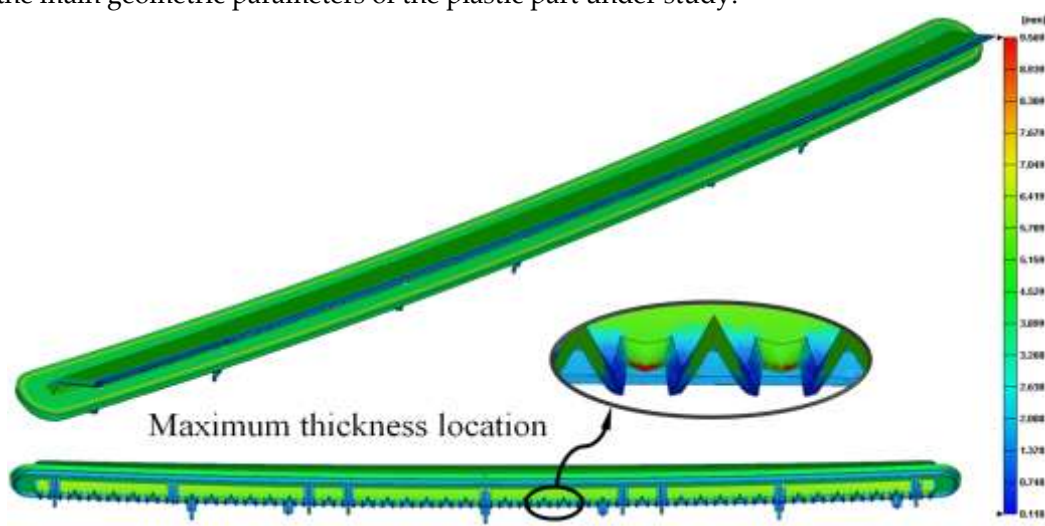

**Fig. 2** Thickness map distribution of the plastic part

**Table. 1** Plastic part main dimensions

| Maximum thickness [mm] | Average thickness [mm] | Width [mm] | Length [mm] | Height [mm] |
|---|---|---|---|---|
| 9.6 | 4.1 | 62.0 | 630.0 | 40.0 |

As can be seen in Table 1 and Fig. 3, the plastic part has a particular topology, features, and geometric parameters, which differ from the usual standards in the plastics industry. The nominal length of the plastic part is 630 mm. The difference between the average and maximum thickness is 5.5 mm. This thickness variation is given by the need to include an upper wing and several collimating elements, to project and direct the light beam coming from the LEDs towards the pillow elements (see Fig. 1). The part is made up of a set of reinforcing elements of variable dimensions that greatly increase the thickness of the part in its lower area, making it difficult to manufacture. These features and geometric parameters directly affect the design and manufacture of the main elements of the cooling system and its functionality.

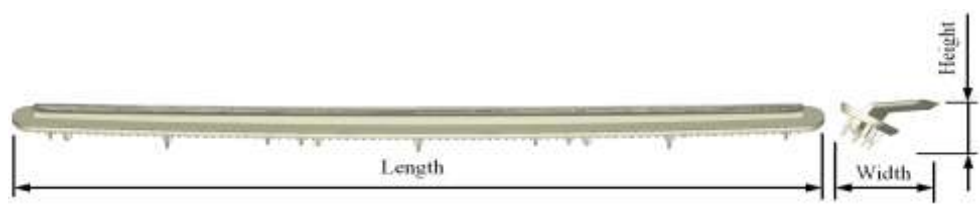

**Fig. 3** Plastic part main dimensions

*2.2.- Theoretical background and analytical study of the cooling phase*

The analytical study of the cooling phase of the plastic part under study focuses on the analysis of two technological parameters: time to reach the ejection temperature and longitudinal warpage. On the one hand, the evaluation of the current time that elapses until the plastic part reaches the ejection temperature facilitates the evaluation of the thermal performance associated with the design of the cooling system and the cooling phase of the plastic part. Meanwhile, the longitudinal warpage in a plastic part after cooling determines its compliance with dimensional tolerances and its functional requirements.

2.2.1.- Time to reach the ejection temperature

The manufacturing cycle of the plastic part is determined by different technological phases: filling, packing, cooling, and ejection. Among all of them, the cooling phase is the most important in the thermoplastic injection molding process representing more than 70% of the total time invested in the manufacturing cycle and the one with the greatest influence on energy consumption. The cooling phase begins with the solidification of the gate. This is followed by the solidification of the molten plastic front, which has completed the mold cavity, cooling down until it reaches a sufficient and suitable temperature to go on to the ejection phase. The physical model used to determine the cooling time is established from the heat exchange by conduction that occurs between the thermoplastic material and the coolant flow. Consequently, the hypothesis is established that the thermal exchange by convection and radiation between the main elements of the injection mold and the external environment is neglected, as this represents less than 5% of the total heat exchange. The heat flow by conduction is defined, according to Eq. 1, using Fourier's differential equation, reduced to one dimension. Although this thermal analysis can be applied to the three main directions of space, the heat exchange flow between the thermoplastic material and the coolant flow is carried out in only one direction, the main direction of the thickness so the unidirectional approach of the analysis is valid [5]. In this way, the one-dimensional heat flow or Fourier equation reduced to one dimension can be expressed as:

$$\frac{\partial T}{\partial t} = \alpha_p \cdot \frac{\partial^2 T}{\partial^2 z}$$

(1)

Being $\alpha_P$ [m²/s] the coefficient for the thermal diffusivity of the mold material, taking into account that after the filling of the mold the temperature of the molten plastic is constant, while the temperature of the surface of the mold changes until it reaches a value stationary. For these conditions, a solution of a particular type can be found (see Eq. 2) for Eq. 1.

At a belonging point to the mold cavity it is possible to analyze the temperature of the molten plastic through the convergence of the Fourier series as indicated in equation 1.

$$T_{z=0}(t) = T_{coolant} + (T_{melt} - T_{coolant}) \cdot \sum_{m=0}^{\infty} \frac{(-1)^m}{2 \cdot m + 1} \cdot e^{\frac{\pi \cdot (2 \cdot m + 1)^2 \cdot \alpha_p}{T_p^2} \cdot t}$$

(2)

Solving Eq. 2 concerning the cooling time variable, the expression that determines the cooling time $t_{cooling}$ is defined in Eq. 3.

$$t_{cooling} = \frac{T_p^2}{\pi^2 \cdot \alpha_p} \cdot Ln\left(\frac{4}{\pi} \cdot \frac{T_{melt} - T_{mold}}{T_{eject} - T_{mold}}\right)$$

(3)

Where $T_P$ [m] represents the maximum thickness of the plastic part, $T_{melt}$ [ºC] represents the temperature of the molten plastic front, $T_{mold}$ [ºC] represents the surface temperature of the injection mold cavity and $T_{eject}$ [ºC] represents the recommended ejection temperature for the thermoplastic material (see Fig. 4).

Also, it should be noted that the boundary conditions established to define Eq.3 are that the thickness of the plastic part is much less than the rest of its dimensions and that the direction of the heat flow, which is exchanged between the plastic part and the coolant, is perpendicular to the melt plastic flow direction [5].

Secondly, the magnitude of the main thermal variables that define Eq. 3 are established and recommended by the manufacturer or supplier of the thermoplastic material. Also, Eq. 3, determined by a physical model, establishes an analytical and approximate cooling time. Well, in practice, the cooling time can be substantially longer if the mold maker considers that the plastic part requires more cooling time to achieve the required functional specifications and geometric tolerances. In addition, the definition of the magnitude of the surface temperature of the mold cavity $T_{mold}$ [ºC] (see Eq. 3) depends directly on the design and dimensioning of the main elements that make up the injection mold cooling system. Therefore, an optimized design adjusted to the features and geometric properties of the plastic part can significantly reduce the result of the cooling time, improving the sustainability of the process.

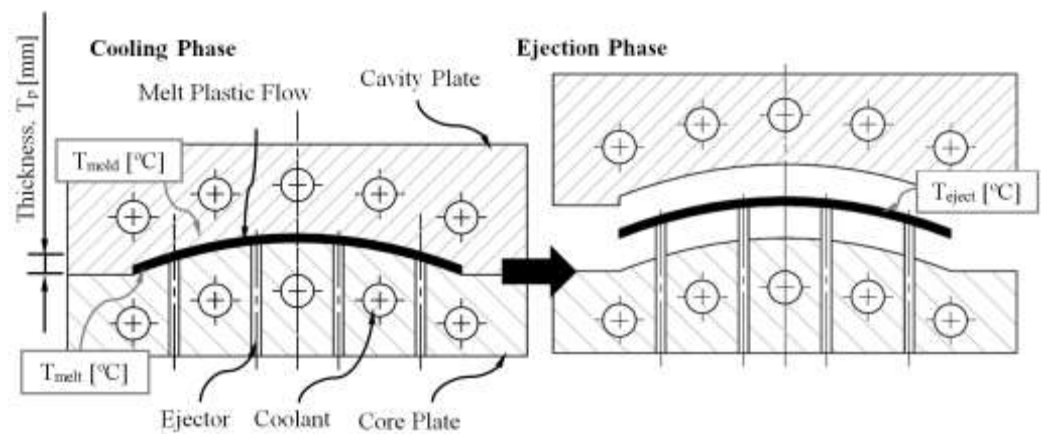

**Fig. 4** Thermal variables representation of the physical model

2.2.2.- Warpage analysis process

The warpage map generated on the plastic part is a result of its manufacturing process. This map defines the final quality of the part, according to its functional requirements and geometric tolerances. In general, the warpage map is related to the volumetric shrinkage that it undergoes during the different phases of the plastic injection cycle. In particular, the main physical effect that generates volumetric shrinkage in the plastic part is the pressure and temperature gradients that develop along its surface, especially after the cooling phase.

According to [48] the differential volumetric shrinkage, caused by temperature gradients and stresses along with the geometry of the plastic part, cause longitudinal warpage outside its main plane (see Fig. 5) and can be expressed according to Eq. 4.

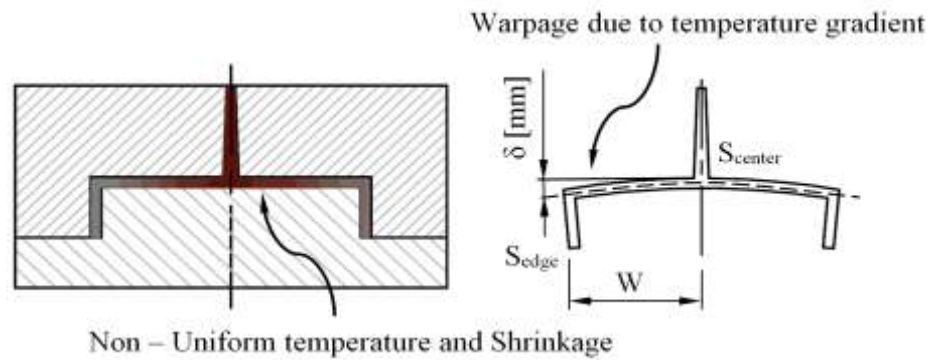

**Fig. 5** Warpage due to differential shrinkage of temperature gradient

$$\delta = \sqrt{W^2 - \{W \cdot [1 - (S_{edge} - S_{center})]\}^2}$$

(4)

Where W [m] represents the separation distance between the center of the plastic part and its outer contour, $S_{edge}$ represents the percentage of linear shrinkage that is obtained along the outer contour of the plastic part, and $S_{center}$ represents the percentage of linear shrinkage obtained in the central region of the plastic piece. Likewise, the volumetric shrinkage parameter is defined from the linear shrinkage experienced by the thermoplastic material in the three main directions of the space. In this way, if the hypothesis is established in which the thermoplastic material is considered to be isotropic, the linear shrinkage that it experiences in the three main directions of space is the same.

And, therefore, the linear shrinkage experienced by the thermoplastic material can be expressed as a function of the magnitude of volumetric shrinkage, according to Eq. 5.

$$S = 1 - \sqrt[3]{r_v}$$

(5)

Where S represents the percentage of linear shrinkage of the thermoplastic material and $r_v$ represents the percentage of volumetric shrinkage of the thermoplastic material. Finally, to complete the analytical study that determines the longitudinal warpages of the plastic part, Tait's double domain equations are used, which characterize the compressibility behavior of the thermoplastic material. From this analytical model, and as shown in Eq. 6, the specific volumes of the thermoplastic material can be defined from its temperature and pressure.

$$v(T,P) = v_o(T) \cdot \left\{1 - 0.0894 \cdot \ln\ln\left[1 + \frac{P}{\beta(T)}\right]\right\}$$

(6)

Where $v_0$ [m³/kg] represents the reference specific volume of the thermoplastic material for the analyzed temperature, β [Pa] represents the compressibility of the thermoplastic material for the analyzed temperature, and P [Pa] the analysis pressure of the thermoplastic material. Furthermore, this set of variables can be expressed according to Eq. 7 and Eq. 8.

$$v_o(T) = b_1 + b_2 \cdot (T - b_5)$$

(7)

$$\beta(T) = b_3 \cdot e^{-b_4 \cdot (T - b_5)}$$

(8)

Where $b_1$ [m³/kg], $b_2$ [m³/kg], $b_3$ [Pa], $b_4$ [1/K] and $b_5$ [K] are technological parameters specific to the thermoplastic material, provided by its manufacturer and supplier. Finally, to define the longitudinal shrinkage (see Eq. 4) suffered by the plastic part, both in its outer contour and in its central region, the percentage of volumetric shrinkage of the thermoplastic material (see Eq. 5) is defined according to Eq. 9. All these requirements, for temperature and pressure conditions between the instant before the beginning of the cooling phase, and the end of the manufacturing process.

$$r_v = \frac{v\ (20^oC\ ,0\ MPa)}{v\ (T_{pack-cooling}, P_{pack-cooling})} \tag{9}$$

*2.3.- Materials*

In this item, we proceed to define the features and properties of the materials used during the research work both for the main elements that make up the injection mold and for the thermoplastic material of the part under study. In this case, the polymer family used to manufacture the plastic geometry is Polymethyl methacrylate (PMMA). This polymer is a highly transparent thermoplastic that is established by the polymerization of the methylmethacrylate monomer. In this way, thanks to its optical and aesthetic features and its high surface strength, this thermoplastic material is used as a lightweight alternative to glass and polycarbonate (PC) when greater transparency, and strength to ultraviolet rays, and high mechanical performance against impacts are required. For this reason, the field of application of this thermoplastic material focuses on the automotive sector, especially for the manufacture of lighting components included in vehicle headlights. In particular, the trade name of the thermoplastic material used is Plexiglas 8N [49]. According to the physical models analyzed in this manuscript, Table 2 shows the mechanical, thermal, and rheological properties of this thermoplastic material according to the information provided by the supplier and manufacturer.

**Table. 2** Magnitude of the main properties of the material Plexiglas 8N

| Nomenclature | Units | Description | Value |
|---|---|---|---|
| $\alpha_P$ | m²/s | Thermal diffusivity | $8.913 \cdot 10^{-8}$ |
| $\varrho_P$ | kg/m³ | Density | 1172.5 |
| $C_P$ | J/kg·°C | Specific heat | 1555.0 |
| $T_{melt}$ | °C | Plastic melt temperature | 235.0 |
| $T_{mold}$ | °C | Mold temperature | 80.0 |
| $T_{eject}$ | °C | Ejection temperature | 94.0 |
| $T_{freeze}$ | °C | Freeze temperature | 132.0 |
| $b_1$ | m³/kg | Tait's dual-domain model coefficient | $0.869 \cdot 10^{-3}$ |
| $b_2$ | m³/kg | Tait's dual-domain model coefficient | $5.679 \cdot 10^{-7}$ |
| $b_3$ | Pa | Tait's dual-domain model coefficient | $1.9492 \cdot 10^{8}$ |
| $b_4$ | 1/K | Tait's dual-domain model coefficient | 0.004633 |
| $b_5$ | K | Tait's dual-domain model coefficient | 394.69 |
| $E_P$ | MPa | Elastic modulus | 3300.0 |
| $\tau_P$ | - | Poisson's ratio | 0.38 |
| CLTE | 1/K | Coefficient of linear thermal expansion | $8.0 \cdot 10^{-5}$ |
| UOI | - | Un – oriented refractive index | 1.49 |
| FSC | cm²/dyne | Flow-induced stress – optical coefficient | $-6.0 \cdot 10^{-11}$ |
| TSC | cm²/dyne | Thermally-induced stress – optical coefficient | $-4.6 \cdot 10^{-13}$ |

Likewise, it should be noted that the manufacturing material can be subjected to a chemical recycling process. In other words, this material, in addition to encountering the requirements and specifications of the plastic part, is recyclable and maintains its original mechanical, thermal, and chemical properties without compromising the sustainability of the manufacturing process.

On the other hand, metallic materials have been defined as the main elements that make up the injection mold systems. For the main insert of the injection mold, a steel alloy called 1.2709 has been used. The selection of this metallic material is justified due to the need to use a 3D additive manufacturing process, based on laser sintering (SLS) technology, which allows the optical fabrication of green conformal cooling channels. In addition, to optimize the cooling of the plastic part under study, it is proposed to use a high conductive steel alloy – 50 at 44 HRC material for the auxiliary elements of the cooling system [50]. The high thermal capacities and performance of this material, together with the location and design of the auxiliary cooling elements, optimize the exchange thermical between part and mold. Especially, in those areas where the effectiveness of the cooling channels decreases and the temperature gradients are greater. Table 3 and Table 4 present several properties of the metallic materials used. The materials properties have been provided by the plastic suppliers.

**Table. 3** Thermal, physical and mechanical properties of Steel alloy 1.2709

| Description | Units | Value |
|---|---|---|
| Density | kg/m$^3$ | 8000 |
| Heat capacity | J/kg·K | $4.62 \cdot 10^3$ |
| Elastic modulus | MPa | $2.36 \cdot 10^5$ |
| Yield stress | MPa | 1016.0 |
| Poisson's ratio | - | 0.30 |
| Coefficient of linear thermal expansion | 1/K | $1.29 \cdot 10^{-5}$ |
| Thermal diffusivity | m$^2$/s | $5.55 \cdot 10^{-6}$ |
| Thermal conductivity | W/m·K | 29 |

**Table. 4** Thermal, physical and mechanical properties of Fastcool – 50 at 44 HRC

| Description | Units | Value |
|---|---|---|
| Density | kg/m$^3$ | 7810 |
| Heat capacity | J/kg·K | $4.70 \cdot 10^3$ |
| Elastic modulus | MPa | $2.07 \cdot 10^5$ |
| Yield strength 0.2% | MPa | 1070.0 |
| Poisson's ratio | - | 0.33 |
| Mechanical resistance | MPa | 1400.0 |
| Elongation | % | 17 |
| Coefficient of linear thermal expansion | 1/K | $1.17 \cdot 10^{-5}$ |
| Thermal diffusivity | m$^2$/s | $1.35 \cdot 10^{-5}$ |
| Thermal conductivity | W/m·K | 50 |

*2.4.- Cooling channels design*

In this section, it is described the design of the devices and channels of the traditional cooling layout that is currently used for the industrial optical part under study. They have also presented the proposals for channels and elements of the cooling layout, based on the concept of green conformal cooling channels (GCCC) and Fastcool inserts, which allow the optimization of the cooling phase and the technological parameters and sustainability-related to the manufacture of the complex part.

In the first place, the geometry and topology of the part have a direct influence on the main parameters of the process using injection molds, such as cycle time, the temperature map of the plastic part, the map of longitudinal warpage and residual stresses of the plastic part after the cooling phase, the thermal performance of the cooling system and the uniformity in the thermal exchange between the plastic part and the coolant flow. So, these parameters have great relevance to the surface and dimensional quality of the plastic part and in the energy required for its manufacturing. As shown in Table 1 and Fig. 2, the plastic piece under study has a maximum thickness in the collimator area of 9.6 mm, while the average thickness in the rest of the geometry is equal to 4.1 mm. That is, the thickness ratio is equal to 2.34:1. Therefore, the manufacture of a plastic part with this thickness variation is a challenge since, as the dimensions and thickness gradients in the plastic part increase, the residual stresses, and longitudinal warpage increase similarly. In addition, the area of the part with greater thickness presents an increase in the accumulation of heat due to a slower cooling process. This variation in the temperature of the part generates an increase in the cooling time and, consequently, a non-uniform shrinkage process. Likewise, the topology of the plastic piece makes it difficult for the cooling channels to access the pillow region (see Fig. 1) due to the reduced dimensions of the pillows. The use of traditional cooling channels in this area reduces the thermal exchange between the plastic material and the coolant flow, as well as their thermal performance.

On the other hand, the geometric design of the cooling system also determines the thermal and energy performance of the cooling phase of the plastic part. In particular, as shown in Fig. 6, the plastic part has a traditional injection mold design based on straight drilled channels. This traditional design must meet dimensional requirements that guarantee the structural integrity of the different elements and systems that make up the injection mold. These requirements are the distance between cooling channels, the distance between cooling channels and elements of the ejection system, and the distance between cooling channels and the surface of the mold cavity. The diameter of the cooling channels is 8 mm. The separation distance between the cooling channels themselves and the surface of the injection mold cavity or any other element thereof is 16 mm, that is, twice the value of the diameter of the cooling channels. In this way, the sizing and structural integrity criteria established by the industry are met, with the minimum safety distance between elements of the injection mold always greater than 10 mm. Fig. 6 and Table 5 show the traditional design and magnitude of the geometric variables of the cooling system, for the two cavities of the injection mold, whose manufacture is carried out using computerized numerical control (CNC) techniques.

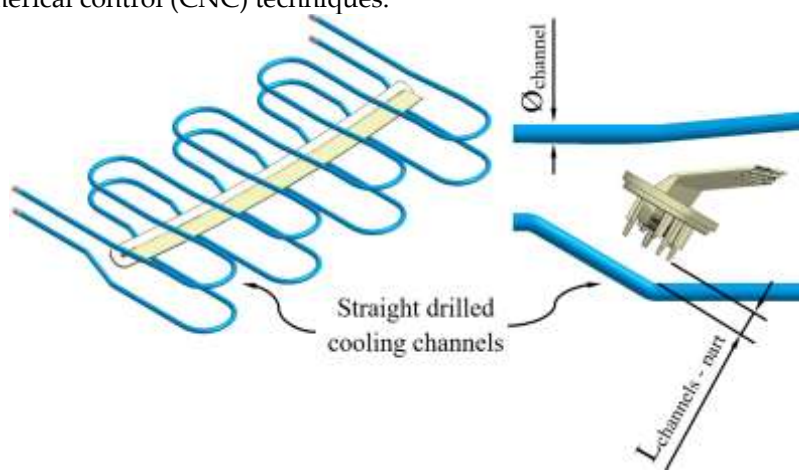

**Fig. 6** Straight drilled cooling channels design

**Table. 5** Geometric variables of the traditional cooling system design

| Description | Units | Value |
|---|---|---|
| Cooling channels diameter | mm | 8.0 |
| Distance between cooling channels and injection | mm | 16.0 |
| Distance between cooling channels and ejection | mm | 16.0 |

Starting from the design and configuration of the straight drilled channels defined for the current cooling system of the plastic part under study and, taking into account the thermal and rheological properties of the thermoplastic material associated with it, using Eq. 3 and Eq. 4, a predictive analytical calculation can be made of the maximum longitudinal warpage obtained after the cooling phase and the total time that elapses during the cooling phase. Table 6 shows the results obtained for said analytical calculation.

**Table. 6** Results obtained from the cooling phase analytical model for the cooling system traditional design

| Description | Units | Value |
|---|---|---|
| Analytical total warpage | mm | 7.1 |
| Analytical cooling time | s | 271.5 |

As can be seen, the analytical results (see Table 6) show that the plastic part under study warps longitudinally by 7.1 mm with a time of the cooling phase equal to 271.5 s, with a traditional cooling system. These analytical results do not meet the functional requirements and geometric and dimensional tolerances established in the industrial sector for optical plastic parts used in vehicle lighting systems. According to the industrial sector, the maximum longitudinal warpage should not exceed 1 mm. Likewise, the resulting cooling time is very long, which implies an increase in cost and energy expenditure associated with the manufacturing process and is contrary to sustainable manufacturing. Therefore, the traditional tools, designs, and technologies for the manufacture and design of the injection mold cooling system are limited and sometimes do not allow the functional requirements and tolerances established by the industrial sector to be met. For this reason, in this manuscript, a new design of conformal cooling channels adapted to the geometry of the plastic part along with the use of auxiliary cooling elements such as Fastcool inserts and the application of new SLS 3D additive manufacturing techniques aim to optimize the cooling phase and the final quality of the plastic part, allowing the functional requirements and tolerances established for the plastic part under study to be achieved.

2.4.1- Green Conformal Cooling Channels Design

The design of the main elements that make up the cooling system of an injection mold focuses on optimizing the dissipation of the heat flow from the plastic part in a uniform manner and in the shortest possible time until it reaches the temperature of ejection. However, this process presents great difficulties and inconveniences in plastic parts with complex surfaces, whose geometric and technological requirements, such as the case study, are impossible to solve with traditional cooling elements or systems.

The conformal cooling channels allow these drawbacks to be solved since they are capable of optimizing the free volume between elements of the injection mold and, thanks to their SLS additive manufacturing process, reducing the distance between the cooling channels and the surface of the plastic part, mainly in geometric areas of difficult access. Likewise, in this manuscript, to improve this problem, a novel design of conformal cooling channels is proposed (see Fig. 7), with a circular section, whose geometric arrangement allows traversing the upper and lower region of the plastic piece, as well as the area concave generated by the support to place the pillows in the correct position to project the light coming from the collimators. On the other hand, although the application of conformal channels for this geometric area is appropriate, sometimes the separation between the channels and the surface of the mold cavity is not enough to dissipate all the heat flow from this region uniformly. As a solution to this design problem, the use of Fastcool inserts in combination with the use of conformal channels is proposed (see Fig. 8 and Fig. 9). Fastcool inserts are characterized by being associated with a metallic material with a high thermal transmission coefficient, which favors heat exchange in areas of the plastic part where high temperature gradients accumulate. However, during the cooling phase, the Fastcool insert exchanges a large amount of heat flow that it is not able to dissipate on its own. The use of conformal cooling channels, in these cases, can be very useful, not so much to directly cool the plastic part but to cool the Fastcool insert itself.

In this way, to evaluate the thermal performance of the application of conformal cooling channels and Fastcool inserts, three different cooling system designs are proposed. A design defined solely by conformal cooling channels, (see Fig. 7) and two hybrid designs that combine conformal cooling channels and Fastcool inserts of different lengths, see Fig. 8 and Fig. 9. Table 7 shows the geometric features of the cooling elements used in the proposed cooling system designs. It should be noted that the geometric parameters that define the proposed cooling systems, conformal and hybrid, have been optimized and adapted to the requirements presented by its own 3D SLS additive manufacturing process, the geometric features of the plastic part under study and the own complexity of the real injection mold with which the present case study is manufactured.

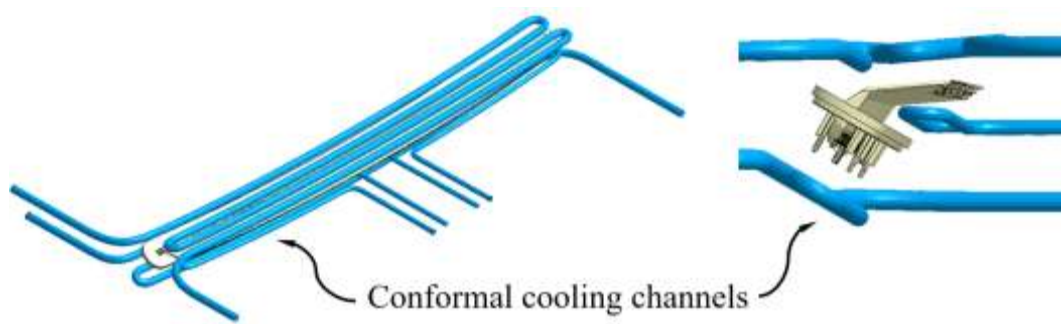

**Fig. 7** Conformal cooling channels design

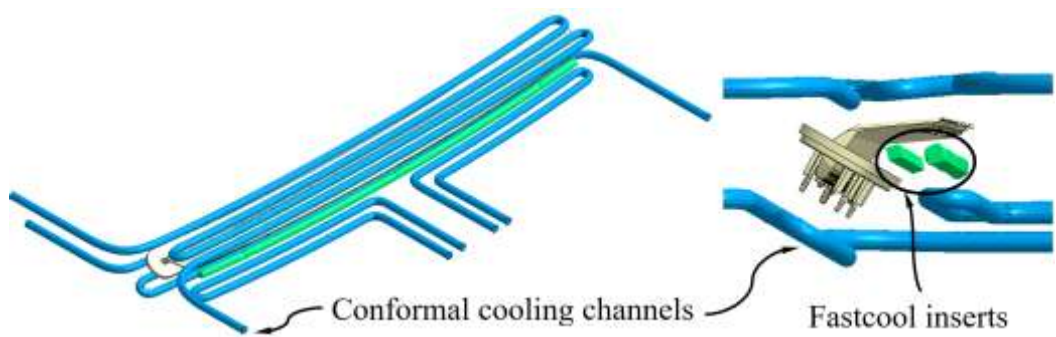

**Fig. 8** Hybrid cooling system design, Fastcool full bars

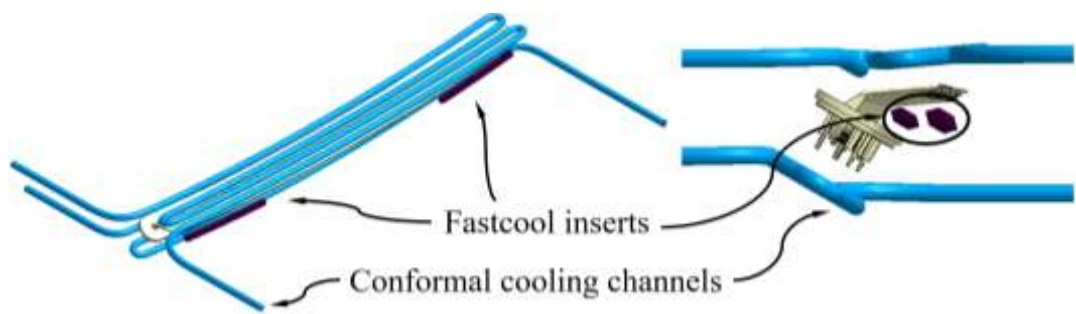

**Fig. 9** Hybrid cooling system design, dashed Fastcool bars

**Table. 7** Geometric variables of the traditional cooling system design

| Description | Conformal cooling system | Hybrid cooling system Fastcool full bars | Hybrid cooling system Dashed fastcool bars |
| --- | --- | --- | --- |
| Cooling channels diameter | 9.0 mm – 6.0 mm | 9.0 mm | 9.0 mm |
| Distance between cooling channels and injection mold cavity | 10.0 mm – 6.0 mm | 10.0 mm | 10.0 mm |
| Fastcool inserts diameter | – | 10.0 mm – 8.0 mm | 10.0 mm – 8.0 mm |
| Distance between Fastcool inserts and injection mold cavity | – | 4.5 mm | 4.5 mm |
| Distance between Fastcool inserts and cooling channels | – | 4.5 mm | 4.5 mm |

Finally, by applying the methodology proposed in this manuscript unlike other cooling proposals, it is possible to minimize the cooling time and improve the surface quality of the plastic part under study, making as the result the manufacturing process sustainable using injection molds.

**3.- Numerical simulations definition for the cooling system thermal modeling**

In this section, the modeling process of the thermal and rheological numerical simulations, carried out to analyze the cooling phase of the plastic part for each of the proposed cooling system designs, is described. In this way, from the numerical results obtained, the thermal performance and the thermal exchange produced between the elements of the cooling system and the plastic piece are evaluated, as well as the temperature gradients, longitudinal warpage, and resulting residual stresses, obtained along the surface of the plastic part. All this is to verify whether the thermal performance of the proposed cooling system designs meets the technological, functional, and dimensional requirements established in the industrial sector for optical plastic parts included in automobile lighting systems. To carry out the different numerical analyzes carried out, the commercial simulation software of the CAE type Moldex3D (R21 version, CoreTech System Co., Ltd., Zhubei City, Taiwan) [50] was used.

The pre-processing and modeling phase of the numerical simulations begins with the meshing operation and assignment of materials for the main computational domains of the injection mold: feeding system (Plexiglas 8N), coolant flow (Water), plastic part (Plexiglas 8N), Fastcool cooling insert (Fastcool 50) and injection mold (Steel alloy 1.2709). The CAE commercial software used in this manuscript has a meshing tool, Moldex Designer, from which the main features of the mesh are established, as well as the type and size of the element used in the discretization of the geometries. Table 8 shows the features and magnitude of the geometric parameters defined for the meshing process. Fig. 10, Fig. 11, and Fig. 12 show the typology of elements used to mesh the main computational domains of the injection mold. These finite elements are of the second-order tetrahedron type (SOLID 186) and have 10 control nodes, which allows the resulting field of temperatures, stresses, and longitudinal warpage to be modeled with greater precision. In addition, to improve the calculation and modeling of the thermal exchange between computational domains, 5 layers of second-order prismatic elements (SOLID 186) of the "Boundary Layer Mesh" type with 15 control nodes are established. As shown in Fig. 10 and Fig. 12, these elements are located on the surfaces of the plastic part and the cooling channels that are in contact with the computational domain of the injection mold. The size of these elements is established by the offset ratio geometric parameter, which is defined as a percentage of the average size of the mesh element of each computational domain to which it belongs.

**Table. 8** Mesh parameters for the standard, conformal, and hybrid cooling systems

| Description | Unit | Value |
|---|---|---|
| Part mesh node count | – | 1,789,045 |
| Part mesh element count | – | 4,494,911 |
| Part mesh volume | cm$^3$ | 269.18 |
| Runner mesh node count | – | 8,435 |
| Runner mesh element count | – | 7,616 |
| Runner mesh volume | cm$^3$ | 0.16 |
| Mesh sizing | mm | 0.75 |
| Element type – Plastic Part | – | Tetrahedral (10 |
| Element type – Boundary Layer | – | Prism (15 nodes) |
| Offset ratio – Boundary Layer | mm | 0.1 |

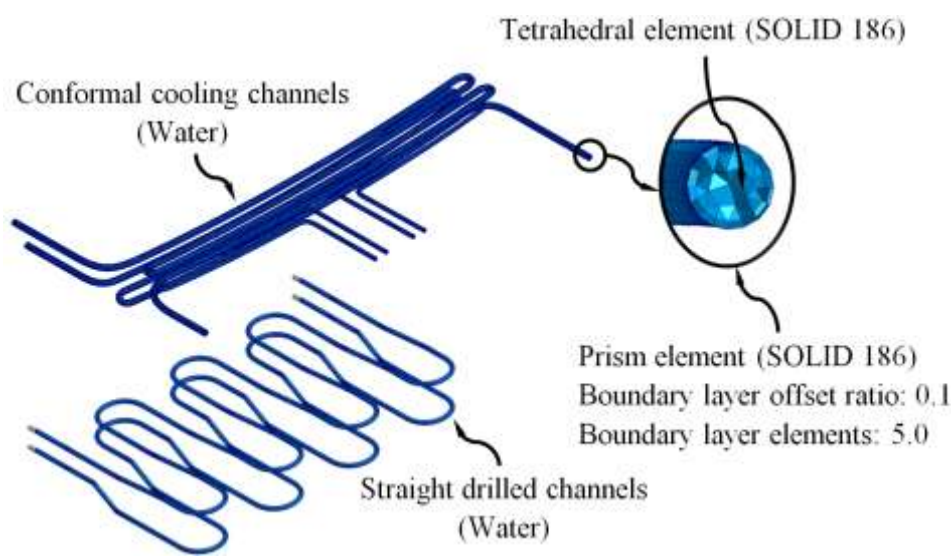

**Fig. 10** Cooling system meshes

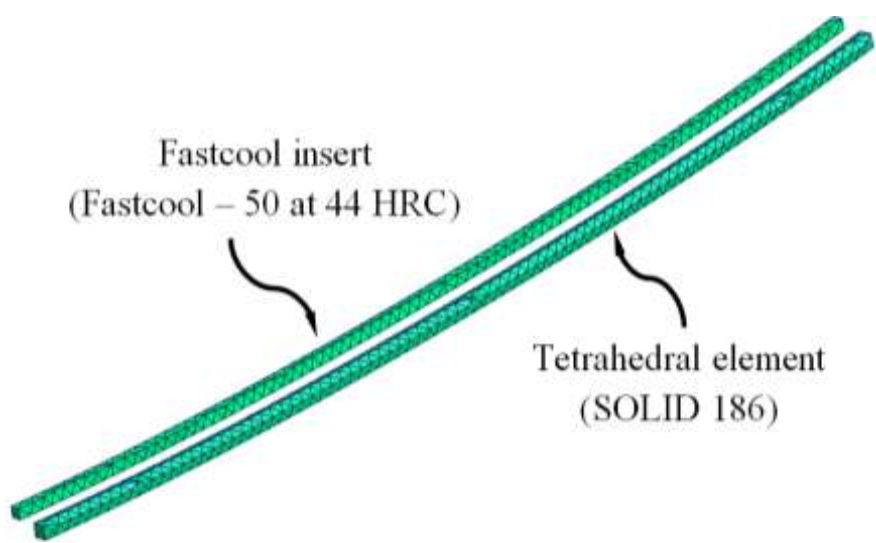

**Fig. 11** Fastcool insert meshes

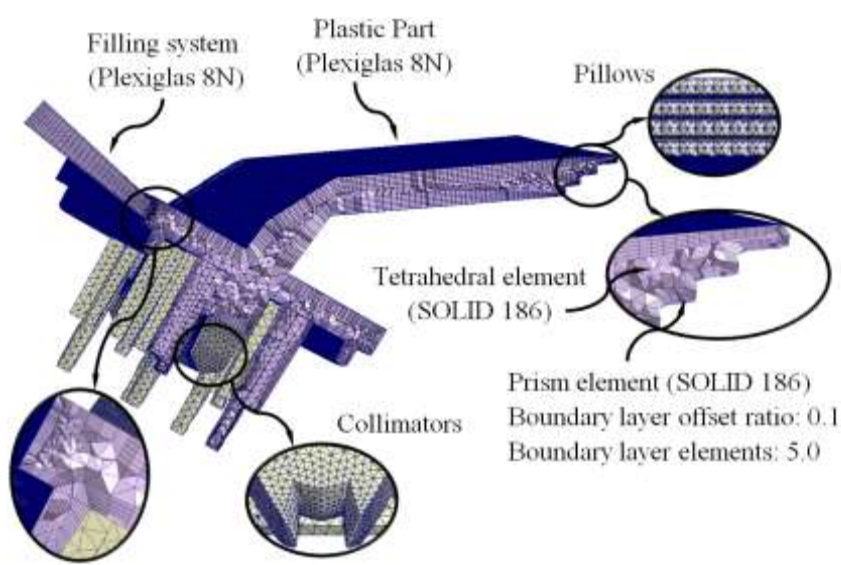

**Fig. 12** Plastic part and filling system meshes

On the other hand, after the meshing phase, the magnitude of the thermal, rheological, and technological parameters of the boundary conditions is established. The definition of these parameters allows simulating each of the different phases in which the manufacturing cycle of the plastic part is divided. In this way, it is possible to model the path of the molten plastic front along with the feeding system and injection mold cavity, the crystallization and solidification of the thermoplastic material, the packing process of the plastic part, and its subsequent cooling. Table 9 shows the magnitude of each of the defined parameters, according to the specifications and recommendations established by the manufacturer and supplier of the thermoplastic material. Furthermore, these technological parameters established during the processing phase of the numerical simulations (see Table 9) are those that can be defined in the real injection machine. Which allows, therefore, to generate a virtual manufacturing model of the plastic piece analogous to the plastic injection process through its corresponding injection mold.

It should be noted that the magnitude of the flow inlet pressure to the cooling system has been established to ensure that it develops in a turbulent flow, that is, that its Reynolds number is greater than $1.5 \cdot 10^4$. In addition, to compare the numerical results obtained for each proposed cooling system design, the flow inlet pressure to the cooling system has been dimensioned so that the Reynolds number is equal to $4.0 \cdot 10^4$ in each numerical simulation.

**Table. 9** Magnitude of the technological parameters defined for the numerical simulations

| Description | Units | Study case – Plexiglas 8N (PMMA) |
|---|---|---|
| Filling time | s | 2.5 |
| Packing time | S | 18.0 |
| VP switch–over (Volume filled) | % | 98 |
| Melt temperature | ºC | 240.0 |
| Mold temperature | ºC | 75.0 |
| Ejection temperature | ºC | 112.0 |
| Coolant temperature | ºC | 75.0 |
| Coolant flow rate | cm³/s | 128.0 (Ø9 mm) |
|  |  | 113.8 (Ø8 mm) |
|  |  | 85.3 (Ø6 mm) |
| Maximum injection pressure | MPa | 140.0 |
| Injection pressure profile | MPa | 140 (0.0s – 2.5s) |
| Maximum packing pressure | MPa | 140.0 |
| Packing pressure profile | MPa | 112 (0.0s – 18.0s) |

The phase aimed at the preprocessing and definition of the numerical simulations requires the establishment of the calculation or computation process following the criteria:
- Transient cooling has been used to analyze the cooling of the plastic part over time.
- An analysis model of the evolution of the dynamic, thermal and physical properties in the cooling channels has been followed according to the configuration: "Run 3D Cooling Channels".
- A solver configuration methodology has been used for each analysis according to a maximum variation of the mold temperature, establishing the values of 1 ºC and 10 ºC as a temperature difference parameter and maximum number of cycles.
- For the turbulence model, it has been established using the roughness parameter in charge of defining the surface that interfaces between the cooling flow and the surface of the channels through which it circulates, this value being 0.02 mm.
- The effects of volumetric shrinkage of the mold, residual stress of the molten plastic front, volumetric shrinkage caused by temperature gradients and longitudinal displacements caused by said gradients have been included.

**4.- Numerical results and discussion**

After completing the pre-processing phase of the thermal and rheological simulations carried out, we proceed to present, evaluate and compare the set of numerical results obtained for each of the cooling system designs proposed. From there, it is possible to determine the configuration of the cooling system that optimizes the cooling phase of the plastic part under study and the one that improves the thermal performance and heat exchange between the plastic part and the coolant flow. First of all, Table 10 and Fig. 13a, Fig. 13b, Fig. 14a and Fig. 14b show the magnitude and distribution along the surface of the plastic part of the resulting parameter "Max. Cooling Time", for each cooling system design analyzed.

Table. 10 Maximum cooling time parameters for each cooling system designs

| Cooling system design | Max. Cooling Time [s] | Reduction [s] | Improvement [%] |
|---|---|---|---|
| Straight drilled channels | 262.550 | – | – |
| Conformal channels and Fastcool full bars | 95.391 | 167.159 | 63.668 |
| Conformal channels and dashed Fastcool bars | 90.578 | 171.972 | 65.501 |
| Conformal channels | 87.427 | 175.123 | 66.701 |

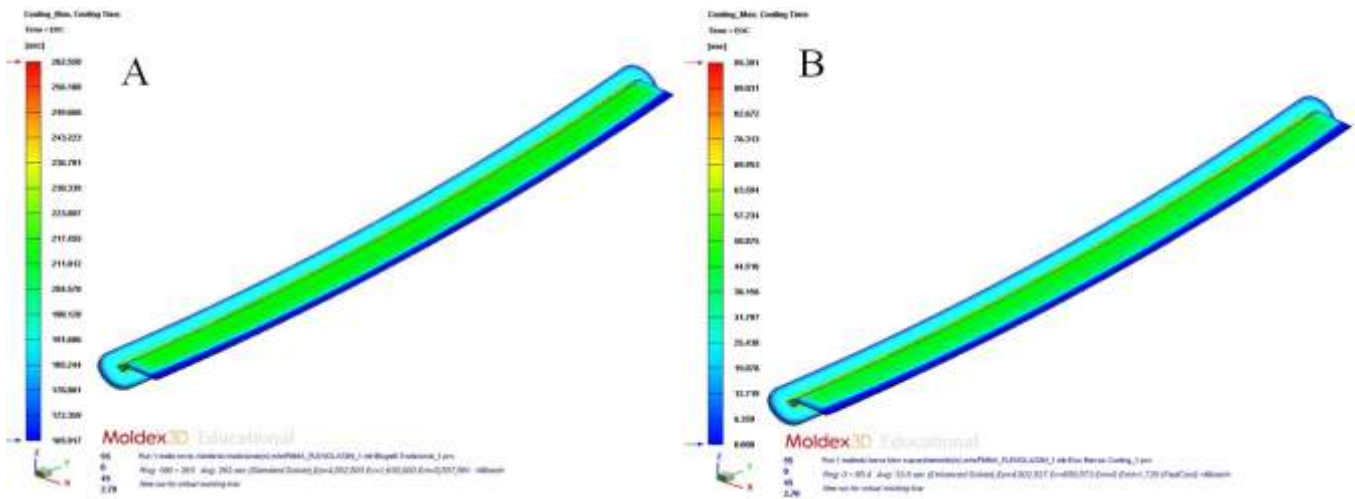

**Fig. 13** Maximum cooling time [s]; Straight drilled channels (A), Conformal channels – Fastcool full bars (B)

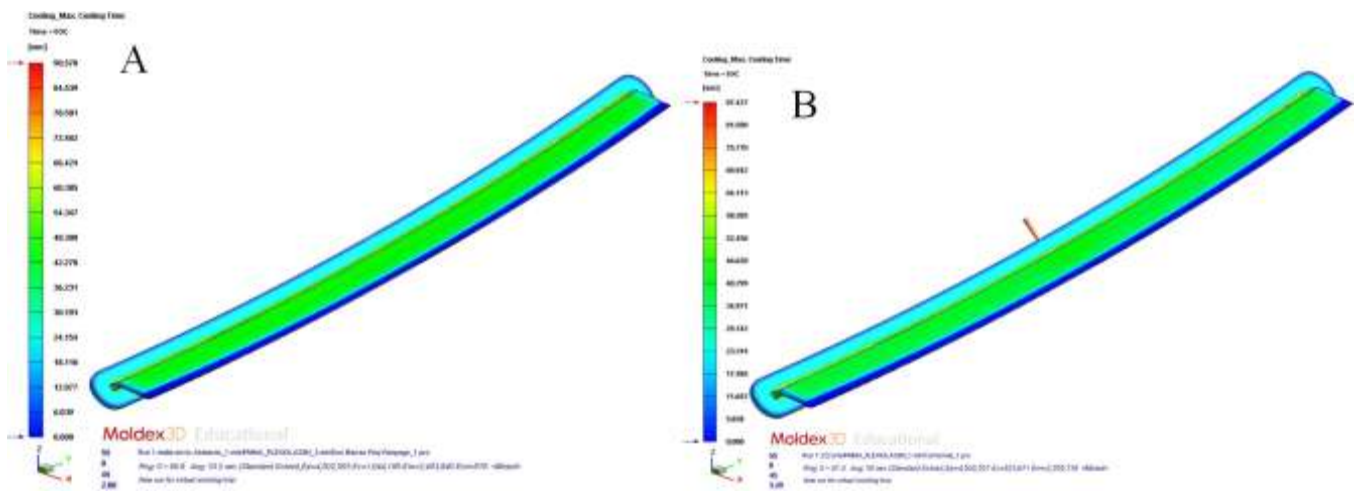

**Fig. 14** Maximum cooling time [s]; Conformal channels – dashed Fastcool bars (A), Conformal channels (B)

From the resulting parameter "Max. Cooling Time", obtained for the design of a traditional cooling system based on straight drilled channels, it is established that the analytical model used to determine the cooling time of the plastic part under study (see Table 6) is valid. According to the analytical model, the cooling time obtained is equal to 271.50 s, while the result obtained in the numerical simulations is equal to 262.55s (see Table 9). In this way, the relative error associated with the calculation of the cooling time parameter (see Eq. 3), obtained through the analytical model presented, is equal to 3.40%, taking as reference the result obtained from the numerical simulations. Therefore, it is determined that the analytical and physical model, proposed in this manuscript to analyze the cooling phase of the plastic part, is admissible and validated, based on the resulting accuracy obtained. Likewise, as shown in Fig. 13a, Fig. 13b, Fig. 14a, and Fig. 14b, the application of conformal cooling channels and auxiliary cooling elements, such as Fastcool inserts, optimize the cooling phase of the plastic part and significantly reduces the time associated with this phase. In particular, according to the results obtained, the application of the proposed cooling system designs improves the cooling time by 167.159 s (Conformal channels and Fastcool full bars, see Fig. 13a), 171.972 s (Conformal channels and dashed Fastcool bars, see Fig. 14a) and 175.123 s (Conformal channels, see Fig. 14b), concerning the result obtained using the traditional cooling system design. This represents an improvement percentage of 63.668% (Conformal channels and Fastcool full bars), 65.501% (Conformal channels and dashed Fastcool bars), and 66.701% (Conformal channels), respectively (see Table 10). In this way, the design based on conformal cooling channels minimizes the magnitude of this variable, compared to the rest of the proposed designs, which in turn implies a significant decrease in the cost and energy expenditure associated with the manufacturing process of the plastic part. Next, Table 11 and Fig. 15a, Fig. 15b, Fig. 16a, and Fig. 16b, show the magnitude and distribution along the surface of the plastic part of the resulting parameter "Mold Temperature Difference", for each design of cooling analyzed.

**Table. 11** Mold temperature difference parameter for each cooling system design

| Cooling system design | Mold temperature | Reduction | Improvement |
|---|---|---|---|
| Straight drilled channels | 23.135 | – | – |
| Conformal channels and Fastcool full bars | 12.757 | 10.378 | 44.858 |
| Conformal channels and dashed Fastcool bars | 10.668 | 12.467 | 53.888 |
| Conformal channels | 4.972 | 18.163 | 78.509 |

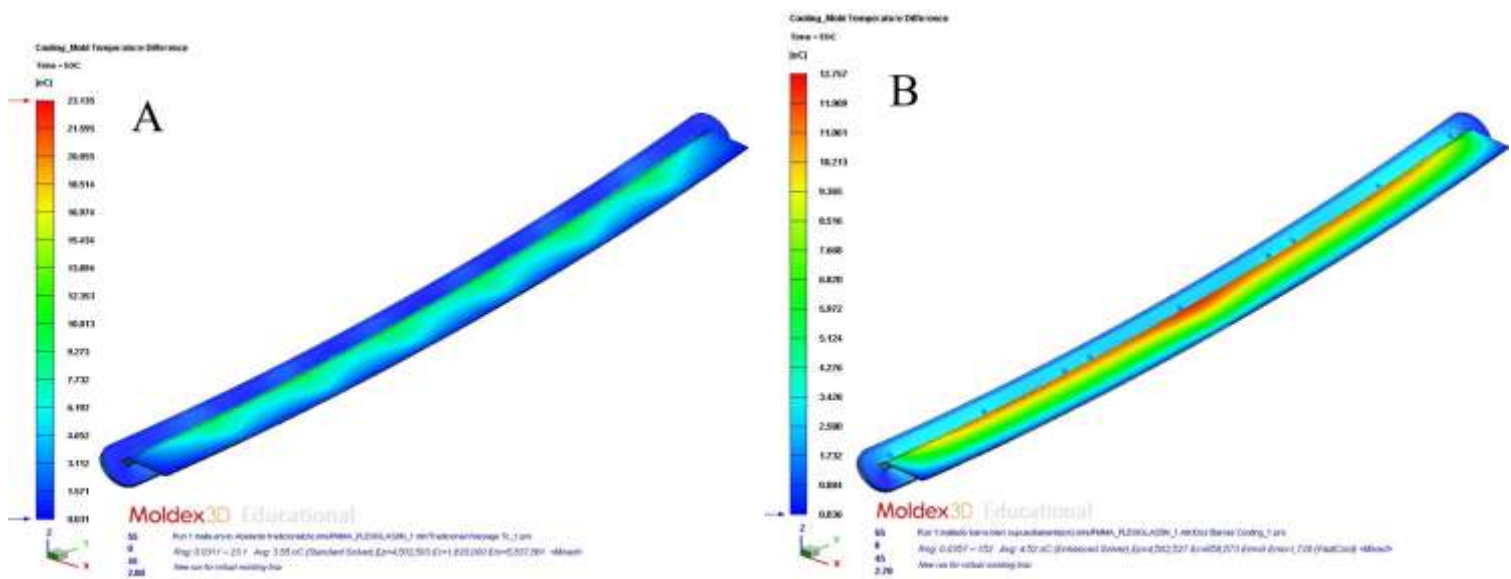

**Fig. 15** Mold temperature difference [ºC]; Straight drilled channels (A), Conformal channels – Fastcool full bars (B)

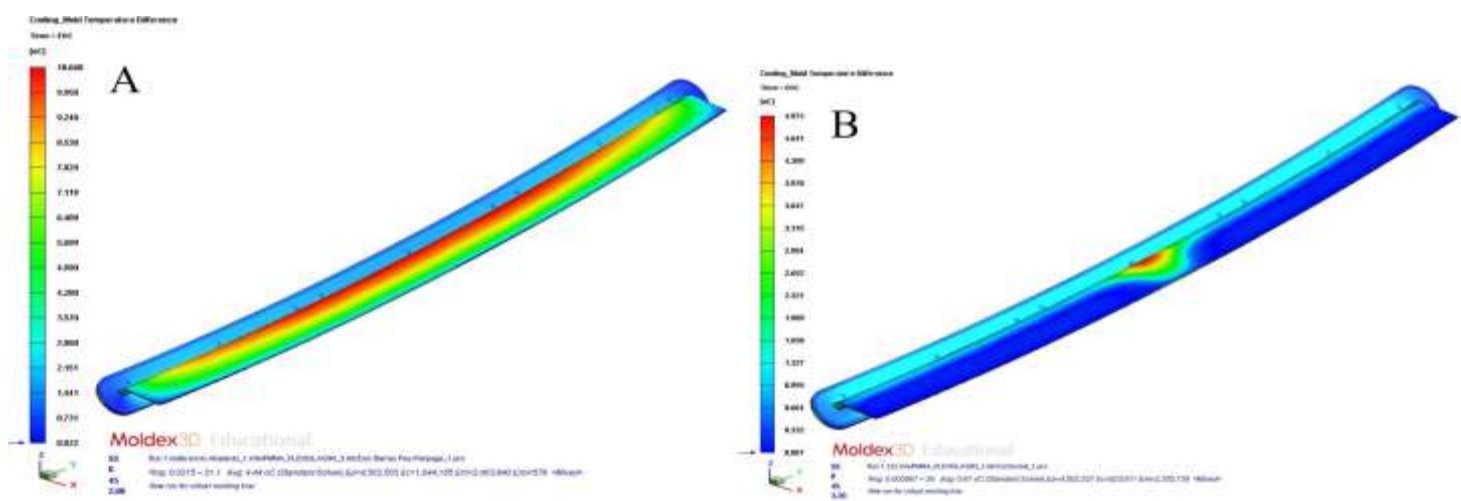

**Fig. 16** Mold temperature difference [ºC]; Conformal channels – dashed Fastcool bars (A), Conformal channels (B)

Based on the results obtained, it is established that the designs of cooling systems with conformal channels and Fastcool inserts improve the uniformity and homogeneity of the temperature map along with the geometry of the plastic part. In this line, the magnitude of the total temperature gradient is notably reduced, which makes it possible to minimize the longitudinal warpage and residual stresses caused by the thermal effects. In particular, according to the results obtained, the application of the proposed cooling system designs improves the temperature gradient to 10,378 ºC (Conformal channels and Fastcool full bars, see Fig. 15a), 12,467 ºC (Conformal channels and dashed Fastcool bars, see Fig. 16a) and 18,163 ºC (Conformal channels, see Fig. 16b), concerning the result obtained with the traditional cooling system design. This represents an improvement percentage of 44.858% (Conformal channels and Fastcool full bars), 53.888% (Conformal channels and dashed Fastcool bars), and 78.509% (Conformal channels), respectively (see Table 11). As can be seen, the design of the conformal channel cooling system optimizes the uniformity of the temperature map and reduces the temperature gradient of the plastic part, compared to designs that include a Fastcool cooling insert. This is because Fastcool inserts, despite being cooled with conformal-type channels, do not exchange heat flow and dissipate it with the same effectiveness as a conformal cooling channel, despite being located closer to the surface of the plastic part (see Fig. 7, Fig. 8 and Fig.9 and Table 7). In other words, in the first moments of the cooling phase, Fastcool inserts quickly exchange the heat flow with the plastic part, but they are not able to dissipate it through the injection mold and the conformal cooling channel increasing its temperature and therefore decreasing its thermal performance. In addition, according to the quality requirements established in the industrial sector of injection plastic molds, it is recommended that the parameter corresponding to the mold temperature difference should be equal to or less than 10ºC. As it shown in Fig. 16a, Fig.16b and Table 11, Conformal channels and Conformal channels – dashed Fastcool bars fullfit this requirement.

Table 12, Table 13 and Fig. 17a, Fig. 17b, Fig. 18a, Fig. 18b, Fig. 19a, Fig. 19b, Fig. 20a, Fig. 20b show the magnitude and distribution along the surface of the plastic part of the resulting parameters "Total Warpage" and "Thermally Induced Residual Stress", for each cooling system design analyzed in this manuscript.

**Table. 12** Total warpage parameters for each cooling system designs

| Cooling system design | Total warpage [mm] | Reduction [mm] | Improvement [%] |
| --- | --- | --- | --- |
| Straight drilled channels | 7.636 | – | – |
| Conformal channels and Fastcool full bars | 1.070 | 6.566 | 85.987 |
| Conformal channels and dashed Fastcool bars | 1.067 | 6.569 | 86.027 |
| Conformal channels | 0.725 | 6.911 | 90.506 |

**Table. 13** Thermally-induced residual stress parameter for each cooling system designs

| Cooling system design | Warpage stress [MPa] | Reduction [MPa] | Improvement [%] |
| --- | --- | --- | --- |
| Straight drilled channels | 48.587 | – | – |
| Conformal channels and Fastcool full bars | 12.743 | 35.844 | 73.773 |
| Conformal channels and dashed Fastcool | 10.219 | 38.368 | 78.968 |
| Conformal channels | 8.803 | 39.784 | 81.882 |

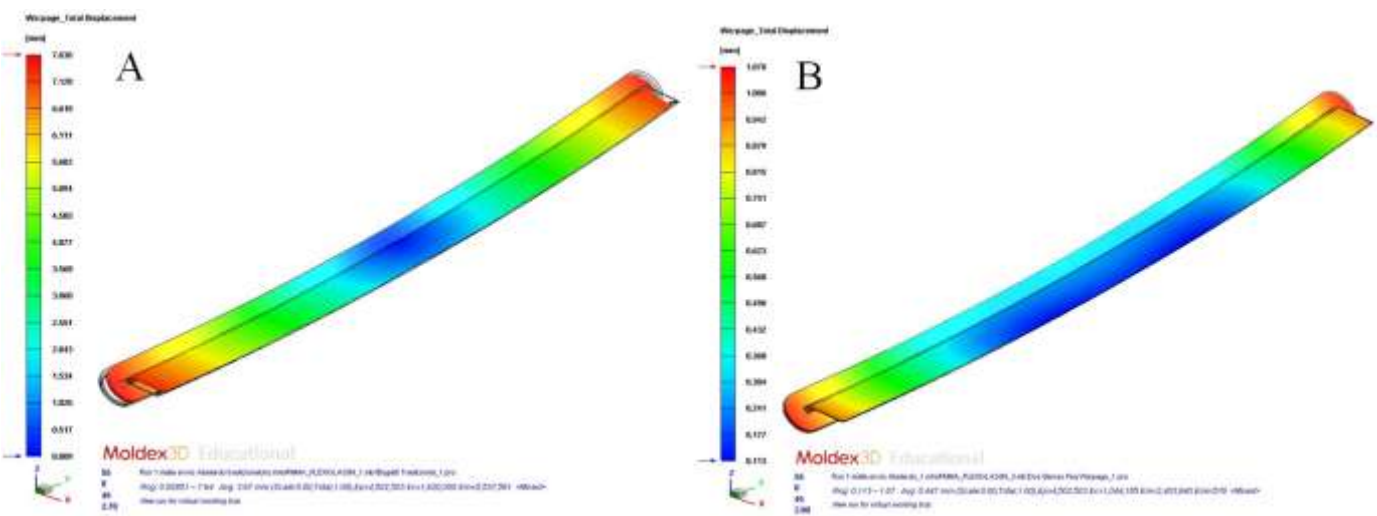

**Fig. 17** Total warpage [mm]; Straight drilled channels (A), Conformal channels – Fastcool full bars (B)

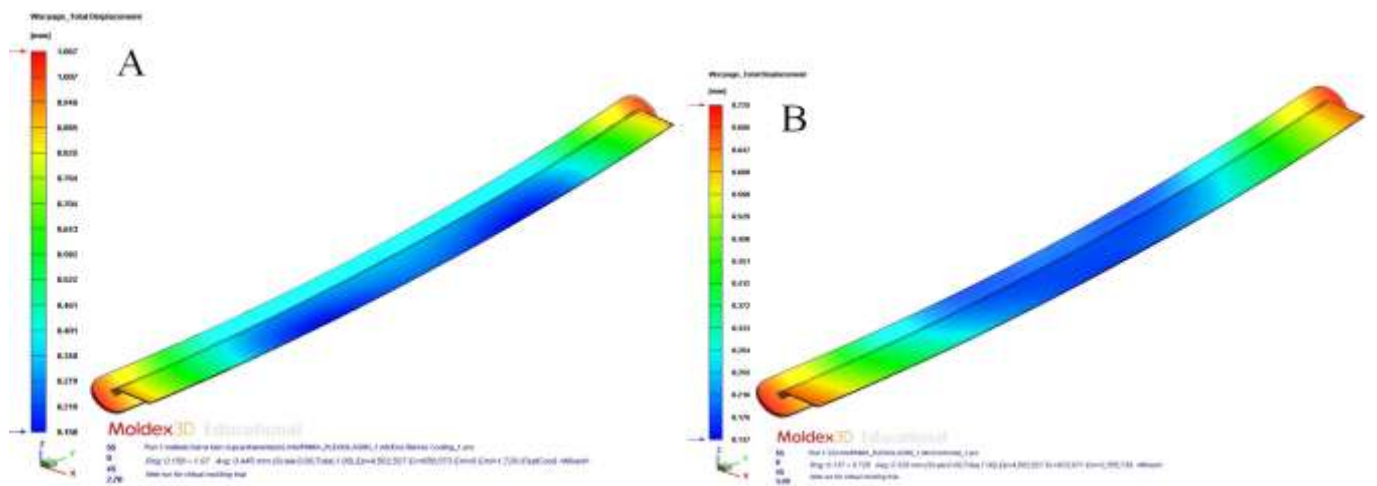

**Fig. 18** Total warpage [mm]; Conformal channels – dashed Fastcool bars (A), Conformal channels (B)

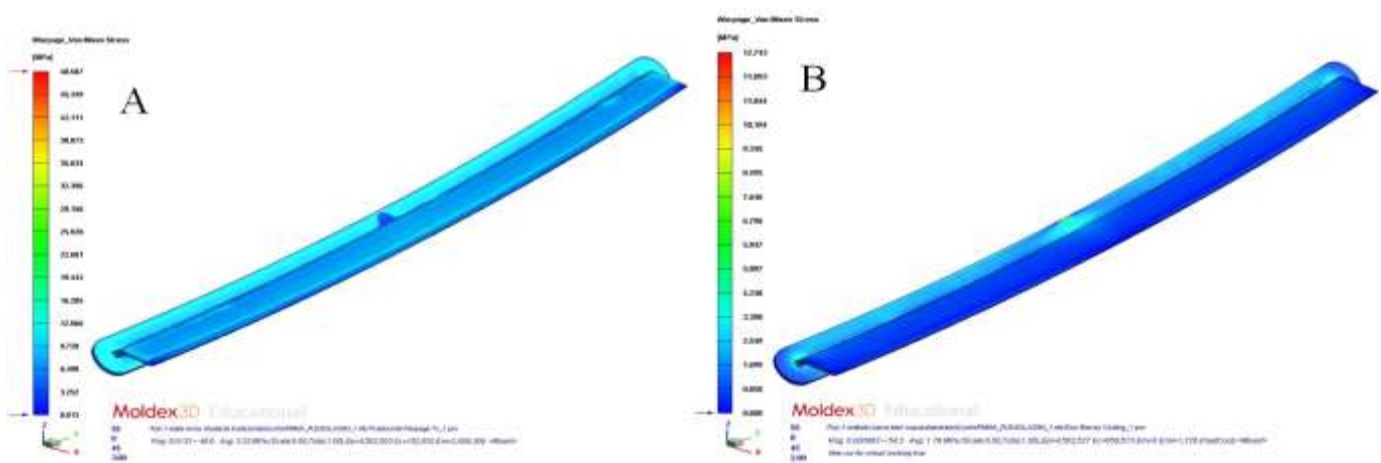

**Fig. 19** Warpage stress [MPa]; Straight drilled channels (A), Conformal channels – Fastcool full bars (B)

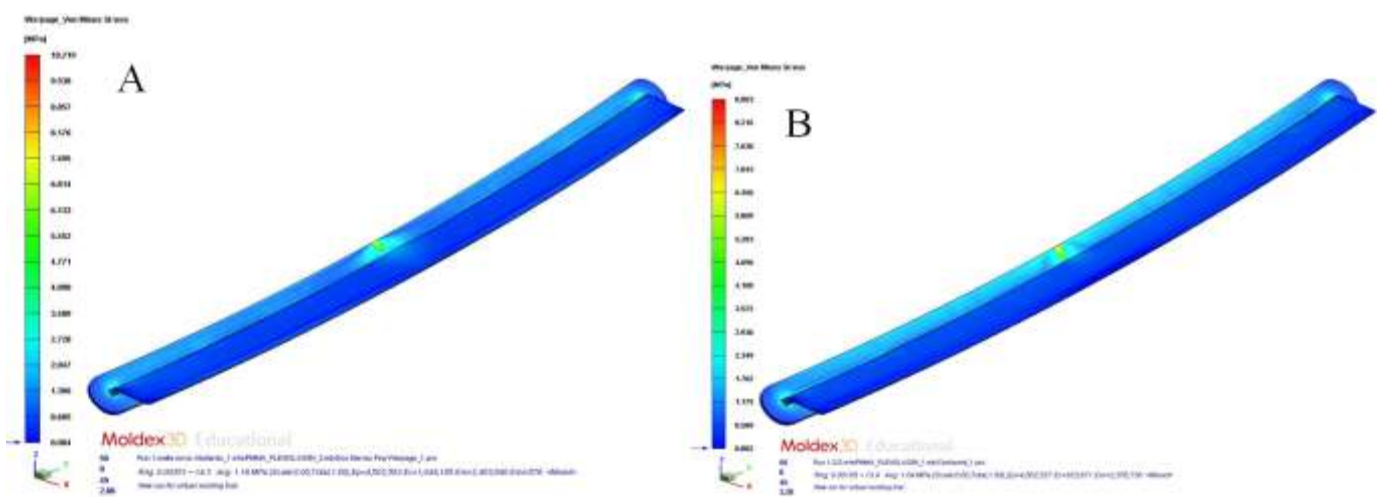

**Fig. 20** Warpage stress [MPa]; Conformal channels – dashed Fastcool bars (A), Conformal channels (B)

The results obtained for the "Total Warpage" and "Thermally Induced Residual Stress" parameters confirm that the application of conformal cooling channels and Fastcool inserts improves the final quality of the plastic part under study. In particular, the resulting longitudinal warpage is reduced by a magnitude of 6.566 mm (Conformal channels and Fastcool full bars, see Fig. 17a), 6.569 mm (Conformal channels and dashed Fastcool bars, see Fig. 18a), and 6.911 mm (Conformal channels, see Fig. 18b), compared to the result obtained with the traditional cooling system design. This represents an improvement percentage of 85.987% (Conformal channels and Fastcool full bars), 86.027% (Conformal channels and dashed Fastcool bars), and 90.506% (Conformal channels), respectively (see Table 12).

In addition, according to the quality requirements established in the industrial sector of injection plastic molds, it is recommended that the parameter corresponding to the mold temperature difference should be equal to or less than 10 °C. As it shown in Fig. 16a, Fig.16b and Table 11, Conformal channels and Conformal channels – dashed Fastcool bars fullfit this requirement.

On the other hand, the residual stresses derived from temperature differences on the surface of the plastic part are also significantly minimized to a magnitude of 35.844 MPa (Conformal channels and Fastcool full bars, see Fig. 19a), 38.368 MPa (Conformal channels and dashed Fastcool bars, see Fig. 20a) and 39.784 MPa (Conformal channels, see Fig. 20b), compared to the result obtained with the traditional cooling system design. This represents an improvement percentage of 73.773% (Conformal channels and Fastcool full bars), 78.968% (Conformal channels and dashed Fastcool bars), and 81.882% (Conformal channels), respectively (see Table 13). As can be seen, the conformal channel cooling system minimizes longitudinal warpage and residual stresses of the plastic part after its manufacturing process. The magnitude of these variables comes, mainly, from the temperature differences along the plastic part. And, as previously defined, conformal cooling channels optimize the uniformity of the temperature map, minimize the temperature gradient of the plastic part and maximize thermal efficiency during the cooling phase, compared to designs that include a Fastcool insert.

Likewise, and analogously to the previously analyzed cooling time parameter, the longitudinal warpage obtained using the physical and analytical model presented (see Table 6) is close to the value obtained using numerical simulations, for the design of a traditional cooling system. Specifically, based on the physical and analytical model presented, the longitudinal warpage obtained for the plastic part under study are equal to 7,356 mm, while the result obtained from the numerical simulations is equal to 7.636 mm. In this way, the relative error associated with the calculation of the longitudinal warpage (see Eq. 4), obtained through the analytical model presented, is equal to 3.667%, taking as reference the result obtained from the numerical simulations. Therefore, it is determined that the analytical and physical model, proposed in this manuscript for the determination of the longitudinal warpage obtained after the manufacturing process of the plastic part, is admissible and validated, based on the resulting precision obtained.

Finally, it should be noted that the design of the cooling system with conformal channels is the only one that meets the functional requirements and the geometric and dimensional tolerances established by the industrial industry for optical plastic parts used in vehicle lighting systems. Well, the longitudinal warpage associated with this design is equal to 0.725 mm and does not exceed the limit value of 1 mm, established by the industrial sector. Well, the longitudinal warpage associated with this design is equal to 0.725 mm and does not exceed the limit value of 1 mm, established by the industrial sector.

### 5.- Conclusions

The manufacture of parts of great length, slenderness, and high thickness ratios for dimensional precision applications is practically impossible using traditional manufacturing means in injection molding processes, respecting the precision and dimensional specifications of the current industry. The warpage in molded plastic parts depends fundamentally on the differential shrinkage caused by temperature gradients along their surface in the cooling process. This fact is accentuated to a greater extent in pieces of great length since in these cases small temperature gradients can give rise to large longitudinal warpage located mainly at the ends of the piece. Warpage influence the number of rejected parts, greatly impacting the sustainability of the production process. Injection molding is focused on large-scale industrial manufacturing, so a poor definition in the design of the tooling of a piece can lead to enormous economic and environmental losses.

The paper presents the use of conformal cooling layouts as a way to reduce cycle time, temperature gradients on the part surface, longitudinal warpage, and residual stresses in the manufacture of highly complex parts with high dimensional requirements. Likewise, the research carried out by the authors analyzes the influence of the use of Fastcool inserts in combination with conformal cooling layouts in reducing warpage in parts with geometry sensitive to this parameter. The results of the research indicate that the use of conformal cooling layouts reduces the cycle time by 175.1 s, 66% below the current cooling time used to manufacture the part through traditional cooling. Concerning the temperature gradient, it improves by 18.16 ºC, being 78.5% better than the current temperature gradient. Finally, the proposal presented by the authors manages to reduce the warpage by 6.9 mm or 90.5% of the current value, achieving final warpage of 0.72 mm, complying with the maximum warpage value required by the industry of 1 mm. The residual stresses have decreased by 39.78 MPa, 81.88% below the current values obtained with traditional cooling. The results validate that the design of the cooling system with conformal channels improves the uniformity of the temperature map of the plastic part, reducing the temperature gradient along the surface, improving the results of the production process, compared to the use of highly conductive inserts of the Fastcool type. This is because Fastcool inserts, despite being cooled with conformal channels, are not capable of exchanging and dissipating the heat flow generated with the same effectiveness as a conformal cooling channel, despite being located much closer to the surface of the plastic part.

Conformal cooling systems are presented as a valid and adequate proposal for the solution of current warpage problems in molded plastic parts where it is highly complex to obtain the dimensional requirements required by the industry. The geometric features for injection molding analyzed in the paper are especially important given that growth in the demand for parts that include these features is estimated [46-47] in the coming years. The results obtained by the authors mark a turning point in the manufacture by molding of this type of highly complex plastic topologies. The results in terms of a decrease in the production cycle time, decrease in carbon emissions, the number of rejected parts, and energy expenditure, place the results of the authors totally in line with the criteria of industrial sustainability demanded by the European Union as well as in the set of industrialized countries.

**Author Contributions**: Investigation, A.T.-A., J.M.M.-C., J.D.D.C.-G., and C.M.-D.; project administration, C.M.-D.; writing—original draft, A.T.-A., J.M.M.-C., and C.M.-D.; writing—review and editing J.M.M.-C. and C.M.-D.; funding acquisition, C.M.-D. All authors have read and agreed to the published version of the manuscript.

**Funding:** This research work was supported by the University of Jaen through the Plan de Apoyo a la Investigación 2021–2022-ACCION1a POAI 2021–2022: TIC-159

**Institutional Review Board Statement:** Not applicable

**Informed Consent Statement:** Not applicable

**Data Availability Statement:** All data included in this study are available upon request by contact with the corresponding author

**Acknowledgments:** Authors acknowledge the support of CORETECH System Co

**Conflict of Interest:** The authors declare that they have no conflict of interest